\documentclass[a4paper]{ifacconf}

\usepackage[T1]{fontenc} 
\usepackage{graphicx}
\usepackage{natbib}
\usepackage{amsmath}
\usepackage{amssymb}
\usepackage{soul,color}
\usepackage{placeins}
\usepackage{tikz}
\usepackage{booktabs}

\usetikzlibrary{arrows.meta,positioning}

\newcommand \aux[1]		{ \hat{#1}}

\newcommand \auxdRelgz	{ \kappa_4 }

\newcommand \auxRelc		{ \hat{\mathcal R}_\mathrm c }
\newcommand \auxRelcz		{ \kappa_1 }
\newcommand \auxRelg		{ \hat{\mathcal R}_\mathrm g }
\newcommand \auxRelgz		{ \kappa_3 }

\newcommand \dzref		{ \dot z_\mathrm{d} }
\newcommand \ddzref		{ \ddot z_\mathrm{d} }
\newcommand \Fmag			{ F_\mathrm{mag} }
\newcommand \Fpas			{ F_\mathrm{pas} }
\newcommand \icoil		{ i }

\newcommand \ksp  { k_\mathrm{s} }

\newcommand \lambdasat	{ \lambda_\mathrm{sat} }
\newcommand \lfrac[2]	{ #1/#2 }
\newcommand \param		{ p }

\newcommand \zmax			{ z_\mathrm{max} }
\newcommand \zmin			{ z_\mathrm{min} }
\newcommand \Rel			{ \mathcal R }

\newcommand \tf			{ t_\mathrm{f} }

\newcommand \vcoil		{ u }
\newcommand \uref	{ u_\mathrm{d} }

\newcommand \zf			{ z_\mathrm{f} }
\newcommand \zref			{ z_\mathrm{d} }
\newcommand \zsp  { z_\mathrm{s} }

\let\originalleft\left
\let\originalright\right
\renewcommand{\left}{\mathopen{}\mathclose\bgroup\originalleft}
\renewcommand{\right}{\aftergroup\egroup\originalright}

\definecolor{bostonuniversityred}{rgb}{0.8, 0.0, 0.0}
\definecolor{limegreen}{rgb}{0.2, 0.8, 0.2}
\definecolor{forestgreen}{rgb}{0.13, 0.55, 0.13}
\definecolor{greenhtml}{rgb}{0.0, 0.5, 0.0}
\definecolor{filln}{rgb}{0.8, 0.9, 1.0}
\definecolor{linen}{rgb}{0.0, 0.1, 0.2}
\definecolor{fillt}{rgb}{1.0, 0.85, 0.7}
\definecolor{linet}{rgb}{0.3, 0.15, 0.0}
\definecolor{filld}{rgb}{1.0, 0.7, 1.0}
\definecolor{lined}{rgb}{0.3, 0.0, 0.15}
\definecolor{fillg}{rgb}{0.77, 0.9, 0.77}
\definecolor{lineg}{rgb}{0.15, 0.2, 0.15}

\begin{document}
\begin{frontmatter}
\hspace{100pt}%
\title{Run-to-Run Adaptive Nonlinear Feedforward Control of\\Electromechanical Switching Devices\thanksref{footnoteinfo}}

\thanks[footnoteinfo]{This work was supported in part via grants PID2021-124137OB-I00, TED2021-130224B-I00, and CPP2021-008938, funded by MCIN/AEI/10.13039/501100011033, by ERDF A way of making Europe, and by the European Union NextGenerationEU/PRTR, in part by grant T45{\_}20R funded by the Government of Arag\'on and in part by the ``Programa Investigo'' funded by the European Union NextGenerationEU.
}

\author[First]{Eduardo Moya-Lasheras}
\author[First]{Edgar Ramirez-Laboreo}
\author[First]{Eloy Serrano-Seco}

\address[First]{Departamento de Informatica e Ingenieria de Sistemas (DIIS) and Instituto de Investigacion en Ingenieria de Aragon (I3A),\\Universidad de Zaragoza, 50018 Zaragoza, Spain,\\ (e-mail:
\{emoya,
ramirlab,
eserranoseco\}%
@unizar.es%
)}

\begin{abstract}
Feedforward control can greatly improve the response time and control accuracy of any mechatronic system. However, in order to compensate for the effects of modeling errors or disturbances, it is imperative that this type of control works in conjunction with some form of feedback. In this paper, we present a new adaptive feedforward control scheme for electromechanical systems in which real-time measurements or estimates of the position and its derivatives are not technically or economically feasible. This is the case, for example, of commercial electromechanical switching devices such as solenoid actuators. Our proposal consists of two blocks: on the one hand, a feedforward controller based on differential flatness theory; on the other, an iterative adaptation law that exploits the repetitive operation of these devices to modify the controller parameters cycle by cycle. As shown, this law can be fed with any available measurement of the system, with the only requirement that it can be processed and converted into an indicator of the performance of any given operation. Simulated and experimental results show that our proposal is effective in dealing with a long-standing control problem in electromechanics: the soft-landing control of electromechanical switching devices.
\end{abstract}


\end{frontmatter}

\section{Introduction}\label{sec:intro}

Feedforward control is widely used for tracking applications because it significantly outperforms other control schemes in terms of response time and tracking accuracy. Since this type of control does not require state information, it is particularly useful for applications in which sensing or estimation of the variables to be controlled is inaccurate or unavailable. Specifically, for differentially flat systems, it is possible to design a feedforward law that provides the control signal from the desired output trajectory without need of solving any differential equation. An advantageous aspect of flatness-based feedforward control (also known as exact feedforward linearization) is that it does not suffer from the well-known robustness problems of its feedback counterpart (exact feedback linearization) due to model parameter uncertainty~\citep{hagenmeyer2003}. In recent years, this type of feedforward control has been proposed for controlling the motion of a wide range of mechatronic systems, such as crane rotators~\citep{bauer2014}, electrical drives~\citep{stumper2015}, electrohydraulic systems~\citep{kim2015}, quadrotors~\citep{greeff2018}, and electrostatic quasi-static microscanners~\citep{schroedter2018}.

Despite the advantages of feedforward controllers, it is well known that a control scheme based solely on a feedforward term, i.e., open loop, is quite sensitive to disturbances and modeling errors. Therefore, it is most usual for feedforward control to be complemented by some form of feedback, as in the previously cited references. This of course implies that the state or output variables can be measured or estimated in real time with sufficient accuracy.

Alternatively, some works propose run-to-run adaptation laws, which are useful for devices under repetitive operation. The key idea of these approaches is that, instead of using frequent measurements of the state or output (which may not even be available), the adaptation law makes use of other auxiliary variables related to the overall control performance of each operation. For example, \cite{blanken2017} has presented recently a unifying framework for run-to-run adaptation of feedforward control based on basis functions. However, this methodology is not applicable to flatness-based feedforward controllers.

Certain mechatronic systems cannot incorporate the required sensors for real-time feedback control due to different reasons, such as economic or space limitations.
Among these devices are switch-type electromechanical devices, such as solenoid valves and electromagnetic relays. 

Thus, for some of these cases in which feedback control is not feasible, run-to-run learning-type adaptation laws have been proposed. However, most of them still require some kind of real-time feedback~\citep{Peterson2004,Benosman2015,Moya-Lasheras2020tmech}, or use model-free input parameterization instead of feedforward controllers~\citep{Yang2013,Mercorelli2012tmech,DiGaeta2015}.

In this paper, we present a new control scheme for electromechanical systems that operate in a repetitive manner. It is particularly suitable for cases in which real-time measurements or estimates of the position are not technically or economically feasible, i.e., systems where real-time feedback control cannot be implemented. It consists of two separate but interconnected blocks. Firstly, a feedforward controller based on differential flatness theory. Thanks to the flatness property---which is satisfied by a large number of electromechanical systems---this block can be parameterized in terms of the physical parameters of the model. Secondly, an iterative adaptation law that exploits the repetitive operation of these devices to modify the parameters of the feedforward block cycle by cycle. As it is shown, this run-to-run law can be fed with any available measurement of the system, with the only requirement that it can be processed and converted into an indicator of the performance of any given operation. In the paper, we describe and apply this method to a long-standing control problem in electromechanics: the soft-landing control of electromechanical switching devices. Simulation and experimental results are presented in order to validate the proposal.

\section{Control-oriented dynamical model}\label{sec:model}

\begin{figure}[t]
	\centering
	\hspace*{\stretch{1}}%
	\includegraphics{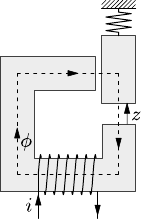} \hfill
	\hspace*{\stretch{2}}%
	\includegraphics{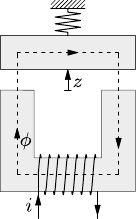} \hfill
	\hspace*{\stretch{2}}%
	\includegraphics{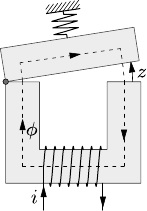} \hfill
	\hspace*{\stretch{1}}%
	\caption{Schematic representation of single-coil reluctance actuators}
	\label{fig:actuators}
\end{figure}

Electromechanical switch-type devices are based on a single-coil reluctance actuator. Schematic diagrams of this kind of actuators with different shapes are represented in Fig.~\ref{fig:actuators}. They all have a fixed core, which is magnetized by a coil current, and a movable core, which is generally attached to a spring or other elastic components. The purpose of the fixed magnetic core is to act as an electromagnet that attracts the movable magnetic core (i.e., armature), closing the gap between them. Given that the single coil is only able to generate magnetic force in one direction, elastic and other passive uncontrollable forces are necessary to move the armature away from the fixed core when the coil current is reduced and the electromagnet is de-energized. In the case of switch-type actuators, the armature position is constrained between two limits.

\FloatBarrier

The system dynamics is as follows. Firstly, the dynamics of the magnetic flux linkage $\lambda$ is derived from the electrical and magnetic equivalent circuits. On the one hand, the coil electrical circuit equation is given by Ohm's law, Faraday's law and Kirchhoff's voltage law, resulting in
\begin{equation}\label{eq:el_circuit}
	\vcoil = R \, \icoil + \dot \lambda, 
\end{equation}
where $\vcoil$, $R$ and $\icoil$ are respectively the coil voltage, internal resistance and current. On the other hand, the current is related to the flux linkage through the magnetic equivalent circuit equation given by Hopkinson's law and Amp\`ere's circuital law,
\begin{equation}\label{eq:mec_circuit}
\icoil = \aux\Rel \, \lambda,
\end{equation}
where $\aux\Rel$ is the magnetic reluctance per turn squared or, equivalently, the inverse of the inductance of the coil. Generally, the reluctance can be defined as a function of the flux linkage (due to magnetic saturation in the core) and the armature position $z$ (due to its dependence on the air gap length in the flux path). A convenient approach is to separate the reluctance into two functions,
\begin{equation}\label{eq:Relgc}
    \hat\Rel =  \auxRelc(\lambda)+\auxRelg(z),
\end{equation}
corresponding to the reluctance contributions of the core, $\auxRelc$, and the gap, $\auxRelg$. Note that there are many possible definitions of these functions, considering different actuators and electromagnetic phenomena. Since their specific expressions are not needed to explain the controller design, in this and the following section we consider them to be arbitrary functions.

Then, the flux linkage dynamics can be derived from~{\eqref{eq:el_circuit}--\eqref{eq:Relgc}}, resulting in the following equation:
\begin{equation}\label{eq:dflux}
	\dot \lambda = - R \, \big( \auxRelc(\lambda)+\auxRelg(z) \big)\,\lambda + u.
\end{equation}

Secondly, the position dynamics during motion is described by Newton's second law,
\begin{equation}\label{eq:newton2}
	m \, \ddot z  =\Fpas(z, \dot z) + \Fmag(z,\lambda),
\end{equation}
where $m$ is the armature mass, and $\Fpas$ and $\Fmag$ are the force functions. The passive forces are encompassed in the function $\Fpas$ which, in a generalized manner, depends on the position (e.g. elastic forces) and the velocity (e.g. viscous friction). The magnetic force, on the other hand, depends on the magnetic flux, which can be indirectly controlled from the coil voltage $u$, as seen in~\eqref{eq:dflux}. More specifically, the magnetic force is given by the function $\Fmag$, defined as follows \citep{ramirez2016new}:
\begin{align}\label{eq:Fmag}
	\Fmag(z,\lambda) = - \frac{1}{2} \, \frac{\partial \auxRelg}{\partial z}\,\lambda^2.
\end{align}

\begin{figure*}[t]
	\centering
	\def\sumoffset{1mm}
	\def\sumoffsetaux{1.5mm}
	\def\nodex{10mm}
	\def\nodey{13mm}
	\def\arrowsep{3mm}
	\def\lwt{0.3mm}
	\def\lwn{0.4mm}
	\def\lwd{0.5mm}
	\begin{tikzpicture}[
		node distance = \nodey and \nodex,
		box/.style = {draw, minimum height=10mm, minimum width=12mm, align=center},
		bigbox/.style = {draw, dashed,draw=linen, minimum height=13mm, minimum width=24mm, align=center},
		sum/.style = {circle, draw, node contents={}},
		>={Stealth[width=2mm,length=3mm]}
		]
		\node (ref) [] {$\zref(t)$};
		\node (ff) [box, fill=fillg, draw=lineg, right=of ref] {Flatness-based\\feedforward controller};
 		\coordinate[right=of ff,xshift=90] (c2);
		\node (traj) [box, fill=filld, draw=lined, left=of ref] {Position\\trajectory design};
		\node (plant) [box, right=of ff,xshift=55mm] {Actuator};
		\coordinate[below=of plant.center, yshift=-0.5*\arrowsep] (c5);
		\coordinate[above=of plant.center] (c7);
		\coordinate[above=of c7] (c8);
		\coordinate[above=of c8] (c9);
		\coordinate[above=of ff.center] (c10);
		\node (hold) [box, fill=filln, draw=linen, above=of ff.center,xshift=40, anchor=center] {Hold};
		\node (cost) [box, fill=filln, draw=linen, above=of plant.center,xshift=-40, anchor=center] {Cost};
		\node (opt) [box, fill=filln, draw=linen, right=of hold,xshift=44, anchor=center] {Run-to-run\\adaptation law};
		\draw[->,draw=lineg,line width=\lwt] (ff) -- node[above] {$\uref^{n}(t)$} (plant);
		\draw[dashed,->,draw=linen,line width=\lwn] (plant) -- node[right] {$y^n(t)$} (c7) -- (cost);
		\draw[dashed,->,draw=linen,line width=\lwn] (cost) -- node[above] {$J^n$} (opt);
		\draw[dashed,->,draw=linen,line width=\lwn] (opt) -- node[above] {$\param^{n+1}$} (hold);
		\draw[dashed,draw=linen,line width=\lwn] (hold) -- node[above] {$\param^{n}$} (c10);
		\draw[dashed,draw=linen,->,line width=\lwn] (c10) -| (ff);
		\draw[draw=lineg,->,line width=\lwt] (ref) -- (ff);
		\draw[dotted,draw=lined,->,line width=\lwd] (traj) -- (ref);
	\end{tikzpicture}
	\caption{Control diagram. The superscript $n$ denotes variables of the $n$th operation. The voltage signal $\uref$ is computed by the feedforward controller as a function of the desired position trajectory $\zref$ and its derivatives. The run-to-run adaptation law uses the operation cost $J$---computed using the system measurable output $y$---to update the parameter vector $p$ of the feedforward controller only once per operation}
	\label{fig:v_ctrl_diag}
\end{figure*}
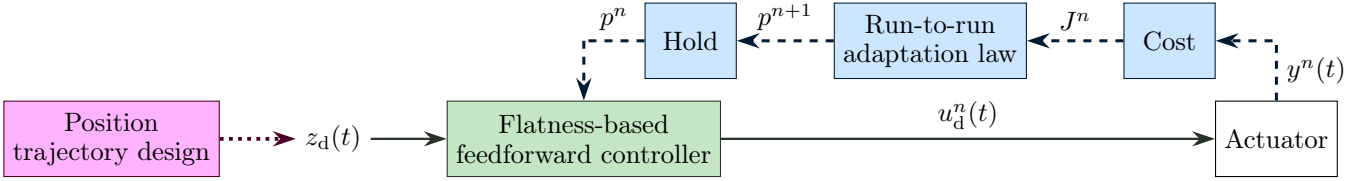

To summarize: the motion dynamics can be described with a state-space representation in which the state variables are the position, $z$, its derivative, $\dot z$, and the flux linkage, $\lambda$; the input is the coil voltage, $u$; and the state dynamics is given by the differential equations~\eqref{eq:dflux} and~\eqref{eq:newton2}. These equations are valid in the domain given by
\begin{equation}
    z \in [\zmin, \; \zmax], \qquad \dot z\in \mathbb R, \qquad \lambda \in (-\lambdasat, \; \lambdasat),
\end{equation}
where $\zmin$ and $\zmax$ correspond to the physical limits of the mechanism and $\lambdasat$ is the maximum flux linkage due to magnetic saturation.

\section{Controller design}\label{sec:ctrl}

The proposed controller is schematized in Fig.~\ref{fig:v_ctrl_diag}. It includes a flatness-based feedforward controller that computes the voltage signal $\uref$ based on the desired position trajectory $\zref$. The voltage signal is fed to the actuator to perform an operation, and a cost or performance index $J$ is calculated. The run-to-run adaptation law adapts the model parameters used in the feedforward controller in order to reduce the cost in the subsequent operations.

\subsection{Trajectory design}\label{sec:zref}

\begin{figure}[t]
	\includegraphics{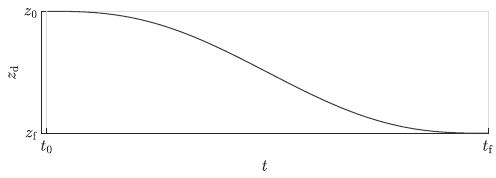}
	\caption{Desired soft-landing trajectory (general form)}
	\label{fig:z_ref}
\end{figure}

Firstly, the desired position trajectory is defined, considering that the objective is to reduce the impact velocities. This paper proposes a soft-landing trajectory with the following boundary conditions:
\begin{equation}\label{eq:bounds}
    \begin{aligned}
	\zref(t_0) &= z_0, & \dzref(t_0) &= 0, & \ddzref(t_0) &= 0, \\
	\zref(\tf) &= \zf, & \dzref(\tf) &= 0, & \ddzref(\tf) &= 0,
    \end{aligned}
\end{equation}
where $t_0$ and $\tf$ are the initial and final user-defined times of the switching operations, and $z_0$ and $\zf$ are the initial and final positions, which respectively correspond to $\zmax$ and $\zmin$ for the closing operations, or to $\zmin$ and $\zmax$ for the opening operations. The trajectory is then designed as a $5$th-degree polynomial, because its six coefficients can be fitted from the six boundary conditions~\eqref{eq:bounds}. The resulting trajectory is represented in a generalized manner in Fig.~\ref{fig:z_ref}.

\FloatBarrier

\subsection{Flatness-based feedforward controller}\label{sec:flatness}

The next step is the design of the feedforward controller, which uses the desired position trajectory to compute a voltage control signal applicable to the actuator. The proposed controller is based on the flatness property of the nonlinear dynamical system. This structural property can be interpreted as an extension of Kalman's controllability of linear systems (a linear system is flat if and only if it is controllable). Proving that a system is flat simplifies greatly the design of many types of controllers. Regarding feedforward control, the flatness property implies that both the state and input variables can be expressed as functions of the output and a finite number of its derivatives~\citep{fliess1995}. This is extremely useful if the variable to be controlled is a flat output, because the input can be calculated without solving any differential equation. Fortunately, the position is a differentially flat output in a large number of electromechanical systems. Thus, we believe that our proposal might be of general use in the design of controllers in the mechatronic field.

It can be easily verified that the position $z$ is a flat output of the presented model: the position and velocity evidently depend on the output and its derivative, whereas the flux and voltage can be expressed as functions of the output and its derivatives by simple manipulations of the model equations. Specifically, the flux can be derived from \eqref{eq:newton2} as
	\begin{equation}
		\lambda = \lambda(z, \dot z, \ddot z) = \pm \sqrt{\frac{2\,\big(\Fpas(z, \dot z) - m \, \ddot z\big)}{\lfrac{\partial \auxRelg}{\partial z}}}.\label{eq:chi_flat}
	\end{equation}
Then, the system input $u$ can be obtained in a similar manner from~\eqref{eq:dflux},
\begin{equation}\label{eq:u_flat}
	\vcoil = \vcoil\left(z, \dot z, \ddot z, \dddot z \right) =
	R\,\left(\auxRelg(z)+\auxRelc(\lambda)\right) \, \lambda + \dot\lambda,
\end{equation}
where \(\lambda\) must be replaced by the expression~\eqref{eq:chi_flat}. The function for $\dot \lambda$ can be obtained by calculating the time derivative of~\eqref{eq:newton2} and solving for $\dot \lambda$.
\begin{multline}\label{eq:dphi_flat}
	\dot \lambda =  \dot \lambda\Big(z,\dot z, \ddot z, \dddot{z}\Big) = \left(\lambda\,\frac{\partial \auxRelg}{\partial z}\right)^{\!-1}\! \Bigg( \frac{\partial \Fpas}{\partial z} \, \dot z + \frac{\partial \Fpas}{\partial \dot z}\, \ddot z\\
    \phantom{=}  
	 - m \dddot{z} - \frac{1}{2} \, \frac{\partial^2\auxRelg}{\partial z^2} \, \dot z \, \lambda^2 \Bigg)
\end{multline}
Note that, for the system to be flat, \(\Fpas\) must be differentiable and \(\auxRelg\) twice differentiable.

The previous equations, which demonstrate the flatness property of the presented system, also serve to design the feedforward control term. In particular, the input signal $\uref$ to achieve the desired trajectory is obtained simply by replacing $z$ with $\zref$ in \eqref{eq:u_flat}.

\subsection{Run-to-run adaptation}\label{sec:r2r}

As already stated, the feedforward controller is very sensitive to modeling errors if it is not accompanied by some form of feedback. Ideally, real-time measurements of the position would be used as feedback to close the loop, but in many cases this is technically or economically unfeasible.
Run-to-run control can be seen as an alternative approach when real-time feedback loops are not an option. It is useful for devices that operate in a repetitive manner, and for which auxiliary measurements can be obtained for evaluating the performance of each operation.

For these reasons, we propose to incorporate a run-to-run adaptation law to the control scheme. Its purpose is to modify the model-based feedforward parameters so as to minimize a certain cost, $J$, which is calculated in each operation. In particular, for the given example, the control objective is to achieve soft-landing trajectories when switching the devices. Thus, the ideal cost would be the absolute value of the impact velocity $v_\mathrm{c}$.
\begin{equation}\label{eq:cost}
    J = \lvert v_\mathrm{c}\rvert
\end{equation}
In practical scenarios where the impact velocities cannot be directly measured or estimated, other concepts related to these but more easily obtainable can be used, e.g. the impact sound or the bouncing duration.

The run-to-run adaptation task can therefore be regarded as a black-box optimization problem. Since the feedforward term is parameterized as a function of the physical parameters of the model, the underlying idea is to find values of these parameters that minimize the cost $J$. In this regard, note that the goal of the adaptation law is not to estimate the model parameters, but to reduce the impact velocity. Thus, there is no guarantee that the adapted values converge to the true ones, but simply to a set of values that minimize the cost. The selected optimization method is the one presented in~\cite{ramirez2017new}. It is based on the Pattern Search method, which is a widely used and studied direct-search optimization method~\citep{lewis2000pattern}. Its convergence properties are directly applicable, so it can be guaranteed that at least a local minimum will be reached provided that the cost function is deterministic.

\section{Simulation Results}

In this section, we present results obtained by simulation, using a specific model presented below. The main reason for performing model-based simulations is to be able to analyze the convergence of the adapted parameters with respect to their nominal values. In other words, the simulations provide a way of testing whether the adaptation law identifies the true parameter values or if,
on the contrary, it only finds a set of values that achieve the proposed control objective.

\subsection{Actuator and simulation model}

The controller presented in the previous section has been derived for a model with arbitrary reluctance functions $\auxRelc(\lambda)$ and $\auxRelg(z)$. In order to perform the simulations, the following expressions have been used:
\begin{align}
    	\auxRelc(\lambda) &= \frac{\auxRelcz}{1-|\lambda|/\kappa_2},\label{eq:relc}\\
    	\auxRelg(z) &= \auxRelgz + \frac{\auxdRelgz \, z}{1 + \kappa_5\, z \, \ln(\kappa_6/z)},   \label{eq:relg}
\end{align}
where the parameters $\kappa_1$, ..., $\kappa_6$ are positive constants. This model, which has been extracted from \cite{Moya-Lasheras2020tmech}, includes magnetic saturation in the core and flux fringing in the air gap. These are the two most significant magnetic phenomena that appear in this class of actuators.

Furthermore, the function $\Fpas$, which encompasses the passive forces, is defined assuming an ideal spring and negligible friction and gravity forces,
\begin{equation}
    \Fpas = - \ksp\,(z - \zsp),
\end{equation}
where $\ksp$ and $\zsp$ are respectively the spring stiffness constant and resting position. 

The nominal model parameters used in the simulations are presented in Table~\ref{tb:param}. 

\begin{table}[t]
    \begin{center}
    \renewcommand{\arraystretch}{1.1} 
    \caption{Model parameter values.}\label{tb:param}
        \begin{tabular}{ccccc}
            Parameter \hspace{0pt} & Value & \hspace{6pt} &
	    	Parameter \hspace{0pt} & Value \\
	    	\cmidrule{1-2} \cmidrule{4-5}
	    	$ \auxRelcz $ & $1.35 \,\mathrm{H^{-1}} $ &&
	    	$ \zmin $ & $0$ \\
	    	$ \kappa_2 $ & $0.0229 \,\mathrm{Wb} $ &&
	    	$ \zmax $ & $10^{-3}\,\mathrm{m}$ \\
    		$ \auxRelgz  $ & $3.88 \,\mathrm{H^{-1}} $ &&
    		$ m $ & $1.6 \times 10^{-3} \,\mathrm{kg} $ \\
    		$ \auxdRelgz $ & $7.67 \,\mathrm{H^{-1}/m} $ &&
    		$ \ksp $ & $ 55 \,\mathrm{N/m} $ \\
    		$ \kappa_5 $ & $1320\,\mathrm{m^{-1}}$ &&
    		$ \zsp $ & $ 0.15 \,\mathrm{m} $ \\
    		$ \kappa_6 $ & $9.73\cdot10^{-3}\,\mathrm{m}$ &&
    		$ R $ & $ 50 \, \mathrm{\Omega}$ \\	
	    	\cmidrule{1-2} \cmidrule{4-5} 
        \end{tabular}
    \end{center}
\end{table}

\subsection{Description of the simulated experiments}

In the simulations, it is assumed that the dynamics of the system is completely described by the model equations previously presented. That is, the model acting in the role of the real system and the feedforward controller are based on exactly the same equations. However, in order to analyze the convergence of the algorithm, it is assumed that there is some uncertainty in the parameter values initially used by the controller. More specifically, the magnetic parameters, $\auxRelcz$, ..., $\kappa_6$, are perturbed following continuous uniform probability distributions whose bounds are $\pm$5\,\% of the nominal values. In essence, these distributions model the errors that might typically be encountered in a parametric estimation process.
On the other hand, the mechanical parameters, i.e., $\zmin$, $\zmax$, $m$, $\ksp$ and $\zsp$, as well as the electrical resistance $R$, are assumed to be perfectly known, as these can usually be measured or estimated with reasonable accuracy. Accordingly, the parameter vector that is iteratively adapted is $p = \left[\auxRelcz\, \cdots\, \kappa_6\right]$, while the rest of the parameters are kept constant.
 
For the sake of brevity, the simulations have been focused on the closing operation, i.e., the motion from $z=\zmax$ to $z=\zmin$. Nonetheless, the control process of the opening operation is completely equivalent. The duration of the soft landing trajectory, $\tf-t_0$, is a parameter that allows to control how aggressive the controller is. In this case it has been set to 3.5 ms, which, for the nominal values of the parameters, ensures that the magnetic flux does not saturate and that the voltage levels are not too demanding.

A total of 10\,000 different experiments have been simulated. For each one of them, the initial parameter vector used by the feedforward controller has been randomly generated according to the rules described above. This way, we analyze the convergence from a large number of initial situations, thereby avoiding incorrect conclusions associated to the results of certain particular cases. Then, the control algorithm has been run for 250 switching operations, searching for a parameter vector that minimizes the impact velocities. As a consequence, a total of 2.5 million switching operations have been simulated, allowing us to deeply analyze the performance of the method.

\subsection{Results and discussion}

\begin{figure}[p]
    \begin{center}
  \includegraphics[]{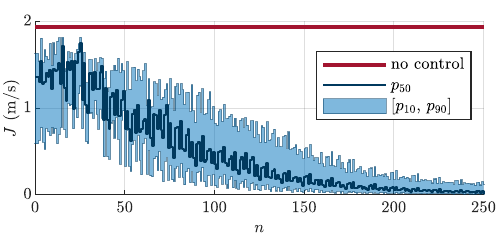}
  \caption{Simulation results. Cost as a function of the number of switching operations. The graph shows the median ($p_{50}$) and the 10th and 90th percentiles ($p_{10}$ and $p_{90}$, respectively) of the distribution of values obtained for the 10\,000 simulated experiments. The cost without control is also represented }
  \label{fig:y_pctl_1}
    \end{center}
\end{figure}

The control results are summarized in Fig.~\ref{fig:y_pctl_1}, which represents the obtained distribution of costs, $J = \lvert v_\mathrm{c}\rvert$, with respect to the switching operation, $n$. To show the control effectiveness, the graph also displays the cost of a conventional switching operation (specifically, with a 30~V constant activation).
As shown, the randomness introduced in the initial parameter vector results in a large variability in the impact velocities of the first switching operations. Despite that, note that the initial impact velocities are in all cases lower than that of the uncontrolled scenario.
Then, the control results improve greatly as the number of iterations increases, which shows the importance of the adaptation law. It takes about 100 switching operations to achieve a cost that is half that of the uncontrolled scenario, and about 200 to make it one tenth. 

\begin{figure}[p]
    \begin{center}
        \includegraphics{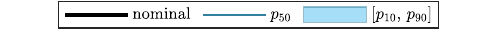}
    	\includegraphics{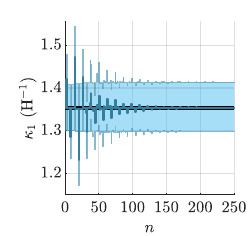} \includegraphics{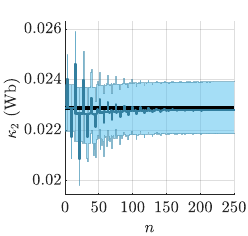}\\\vspace{\floatsep}
        \includegraphics{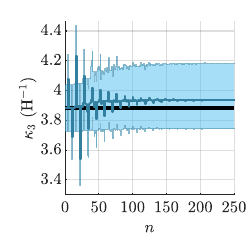} \includegraphics{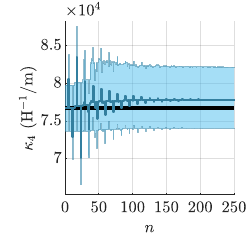}\\\vspace{\floatsep}
        \includegraphics{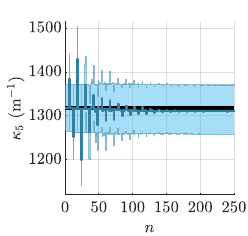} \includegraphics{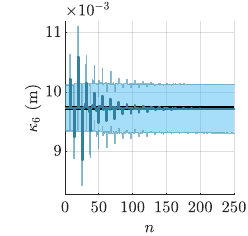}\\
        \caption{Simulation results. Parameter values as a function of the number of switching operations. The graphs show the median ($p_{50}$) and the 10th and 90th percentiles ($p_{10}$ and $p_{90}$, respectively) of the distribution of values obtained for the 10\,000 simulated experiments. The nominal values are also represented }
        \label{fig:p_pctl_1}
    \end{center}
\end{figure}

As explained, it is the parameter adaptation law which causes the impact velocities to be reduced over the course of the operations. The evolution of the parameter values as a function of the number of switching operations is represented in Fig.~\ref{fig:p_pctl_1}. This graph shows the distribution of values of the six parameters that are modified by the adaptive law, as well as their nominal values. It can be seen that the parameters are initially in the $\pm$5\,\% range defined earlier. Then, for each simulated experiment, the adaptation law searches the space of parameters until it reaches virtually stable values from iteration 200 onward. The point to note is that the parameters do not converge to their nominal values, but instead each experiment results in a different final set of parameters. In this sense, our run-to-run adaptive law behaves like most real-time adaptive techniques. That is, the algorithm manages to control the system (in this case, to minimize the impact velocities) as if the true parameters were known. However, it does not guarantee that the adapted parameters converge to the true values. This distinction is important because, as already mentioned, our algorithm should not be seen as a parametric estimation method.


\section{Experimental Results}

Lastly, experimental tests have been also performed to validate the proposal in a real system. In particular, the control has been applied to ten commercial single-pole double-throw power relays of the same family. Ten experiments have been performed with each relay, resulting in a total of a hundred trials. Note that these devices are based on a small switching actuator and, thus, they suffer from the aforementioned problems, i.e., impacts, noise and premature failure. The control objective is thus identical to the simulations: to achieve a soft landing trajectory.
However, given that the impact velocities cannot be easily measured or estimated in these devices, an alternative cost is calculated using an audio signal from a low-cost microphone placed near the device. The key is to realize that the higher the impact velocity, the greater the amount of sound generated in the switching. More specifically, let $v_\mathrm{mic}$ be the voltage signal generated by the microphone. Then, the performance index of a given operation can be computed as
\begin{equation}\label{eq:cost_exp}
    J = \int_{t_\mathrm{0}}^{t_\mathrm{f}+\Delta t} {v_\mathrm{mic}}^2(t) \, \mathrm d t,
\end{equation}
where $\Delta t$ is large enough to capture all the acoustic noise generated during (and after) the switching.

\begin{figure}[t]
    \begin{center}
      \includegraphics[]{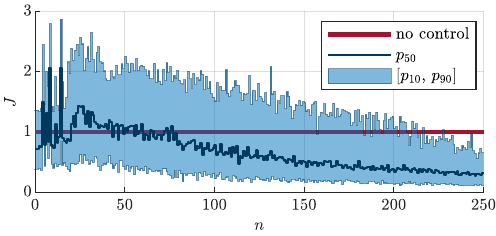}
      \caption{Experimental results. Cost as a function of the number of switching operations. The graph shows the median ($p_{50}$) and the 10th and 90th percentiles ($p_{10}$ and $p_{90}$, respectively) of the distribution of values obtained for the 100 real experiments. The cost without control is also represented }
      \label{fig:y_prctl_exp}
    \end{center}
\end{figure}

The experimental results are presented in Fig.~\ref{fig:y_prctl_exp}. For clarity, the control costs are normalized with respect to the cost of an uncontrolled scenario. Similarly to the simulated case, it can be seen that the initial costs vary significantly. However, all the experiments have used initially the same controller parameters, so in this case the variability must be due to differences between the ten relays or variations in environmental conditions.
Then, the adaptation law manages to reduce the costs as the number of operations increases. Note that, although the convergence is slower than in the simulated experiments, all relays behave better than in the uncontrolled case after the 250 operations. Therefore, we can conclude that our proposal is also effective in a practical scenario.

\FloatBarrier

\section{Conclusion}

In this paper, we have presented a new feedforward control scheme for electromechanical systems in which real-time measurements or estimates are difficult, expensive or simply impossible to obtain. Therefore, the fundamental difference with other adaptive or feedforward methods in the literature is that it does not use real-time feedback. Instead, it includes a run-to-run adaptation law that can use virtually any type of measurement. As has been shown through simulations and experiments, the only necessary condition is that these measurements can be processed to obtain a performance index for any given operation.

As an application example, the scheme has been applied to the soft-landing control of electromechanical switching devices. However, the proposal is sufficiently versatile to be applied to nearly any mechatronic system that performs a repetitive task. Future work will be focused on improving the convergence speed of the method, for example by modifying the search algorithm or by applying dimensionality reduction techniques to the parameter set.


\begin{thebibliography}{17}
\providecommand{\natexlab}[1]{#1}
\providecommand{\url}[1]{\texttt{#1}}
\providecommand{\urlprefix}{URL }
\expandafter\ifx\csname urlstyle\endcsname\relax
  \providecommand{\doi}[1]{doi:\discretionary{}{}{}#1}\else
  \providecommand{\doi}{doi:\discretionary{}{}{}\begingroup
  \urlstyle{rm}\Url}\fi

\bibitem[{Bauer et~al.(2014)Bauer, Schaper, Schneider, and Sawodny}]{bauer2014}
Bauer, D., Schaper, U., Schneider, K., and Sawodny, O. (2014).
\newblock Observer design and flatness-based feedforward control with model
  predictive trajectory planning of a crane rotator.
\newblock In \emph{Amer. Control Conf.}, 4020--4025.

\bibitem[{Benosman and At{\i}n{\c{c}}(2015)}]{Benosman2015}
Benosman, M. and At{\i}n{\c{c}}, G.M. (2015).
\newblock {Extremum seeking-based adaptive control for electromagnetic
  actuators}.
\newblock \emph{Int. J. Control}, 88(3), 517--530.

\bibitem[{Blanken et~al.(2017)Blanken, Boeren, Bruijnen, and
  Oomen}]{blanken2017}
Blanken, L., Boeren, F., Bruijnen, D., and Oomen, T. (2017).
\newblock Batch-to-{{Batch Rational Feedforward Control}}: {{From Iterative
  Learning}} to {{Identification Approaches}}, {{With Application}} to a
  {{Wafer Stage}}.
\newblock \emph{IEEE/ASME Trans. Mechatronics}, 22(2), 826--837.

\bibitem[{Di~Gaeta et~al.(2015)Di~Gaeta, Hoyos~Velasco, and
  Montanaro}]{DiGaeta2015}
Di~Gaeta, A., Hoyos~Velasco, C.I., and Montanaro, U. (2015).
\newblock Cycle-by-cycle adaptive force compensation for the soft-landing
  control of an electro-mechanical engine valve actuator.
\newblock \emph{Asian J. Control}, 17(5), 1707--1724.

\bibitem[{Fliess et~al.(1995)Fliess, L{\'e}vine, Martin, and
  Rouchon}]{fliess1995}
Fliess, M., L{\'e}vine, J., Martin, P., and Rouchon, P. (1995).
\newblock Flatness and defect of non-linear systems: Introductory theory and
  examples.
\newblock \emph{Int. J. Control}, 61(6), 1327--1361.

\bibitem[{Greeff and Schoellig(2018)}]{greeff2018}
Greeff, M. and Schoellig, A.P. (2018).
\newblock Flatness-{{Based Model Predictive Control}} for {{Quadrotor
  Trajectory Tracking}}.
\newblock In \emph{{{IEEE}}/{{RSJ} Int. Conf. Intell. Robots Syst. ({IROS})}},
  6740--6745.

\bibitem[{Hagenmeyer and Delaleau(2003)}]{hagenmeyer2003}
Hagenmeyer, V. and Delaleau, E. (2003).
\newblock Exact feedforward linearization based on differential flatness.
\newblock \emph{Int. J. Control}, 76(6), 537--556.

\bibitem[{Kim et~al.(2015)Kim, Won, and Tomizuka}]{kim2015}
Kim, W., Won, D., and Tomizuka, M. (2015).
\newblock Flatness-{{Based Nonlinear Control}} for {{Position Tracking}} of
  {{Electrohydraulic Systems}}.
\newblock \emph{IEEE/ASME Trans. Mechatronics}, 20(1), 197--206.

\bibitem[{Lewis and Torczon(2000)}]{lewis2000pattern}
Lewis, R.M. and Torczon, V. (2000).
\newblock Pattern search methods for linearly constrained minimization.
\newblock \emph{SIAM J. Optimization}, 10(3), 917--941.

\bibitem[{Mercorelli(2012)}]{Mercorelli2012tmech}
Mercorelli, P. (2012).
\newblock An {{Antisaturating Adaptive Preaction}} and a {{Slide Surface}} to
  {{Achieve Soft Landing Control}} for {{Electromagnetic Actuators}}.
\newblock \emph{IEEE/ASME Trans. Mechatronics}, 17(1), 76--85.

\bibitem[{{Moya-Lasheras} and Sagues(2020)}]{Moya-Lasheras2020tmech}
{Moya-Lasheras}, E. and Sagues, C. (2020).
\newblock Run-to-{{Run Control With Bayesian Optimization}} for {{Soft
  Landing}} of {{Short-Stroke Reluctance Actuators}}.
\newblock \emph{IEEE/ASME Trans. Mechatronics}, 25(6), 2645--2656.

\bibitem[{Peterson and Stefanopoulou(2004)}]{Peterson2004}
Peterson, K.S. and Stefanopoulou, A.G. (2004).
\newblock Extremum seeking control for soft landing of an electromechanical
  valve actuator.
\newblock \emph{Automatica}, 40(6), 1063--1069.

\bibitem[{Ramirez-Laboreo et~al.(2016)Ramirez-Laboreo, Sagues, and
  Llorente}]{ramirez2016new}
Ramirez-Laboreo, E., Sagues, C., and Llorente, S. (2016).
\newblock A new model of electromechanical relays for predicting the motion and
  electromagnetic dynamics.
\newblock \emph{IEEE Trans. Ind. Appl.}, 52(3), 2545--2553.

\bibitem[{Ramirez-Laboreo et~al.(2017)Ramirez-Laboreo, Sagues, and
  Llorente}]{ramirez2017new}
Ramirez-Laboreo, E., Sagues, C., and Llorente, S. (2017).
\newblock A new run-to-run approach for reducing contact bounce in
  electromagnetic switches.
\newblock \emph{IEEE Trans. Ind. Electron.}, 64(1), 535--543.

\bibitem[{Schroedter et~al.(2018)Schroedter, Roth, Janschek, and
  Sandner}]{schroedter2018}
Schroedter, R., Roth, M., Janschek, K., and Sandner, T. (2018).
\newblock Flatness-based open-loop and closed-loop control for electrostatic
  quasi-static microscanners using jerk-limited trajectory design.
\newblock \emph{Mechatronics}, 56, 318--331.

\bibitem[{Stumper et~al.(2015)Stumper, Hagenmeyer, Kuehl, and
  Kennel}]{stumper2015}
Stumper, J.F., Hagenmeyer, V., Kuehl, S., and Kennel, R. (2015).
\newblock Deadbeat {{Control}} for {{Electrical Drives}}: {{A Robust}} and
  {{Performant Design Based}} on {{Differential Flatness}}.
\newblock \emph{IEEE Trans. Power Electron.}, 30(8), 4585--4596.

\bibitem[{Yang et~al.(2013)Yang, Liu, Ye, Chen, and Lu}]{Yang2013}
Yang, Y.P., Liu, J.J., Ye, D.H., Chen, Y.R., and Lu, P.H. (2013).
\newblock Multiobjective optimal design and soft landing control of an
  electromagnetic valve actuator for a camless engine.
\newblock \emph{IEEE/ASME Trans. Mechatronics}, 18(3), 963--972.

\end{thebibliography}
\end{document}